\documentclass[showpacs,pre,twocolumn,numbers,showkeys,superscriptaddress]{revtex4-1}

\usepackage[T1]{fontenc}
\usepackage[utf8]{inputenc}
\usepackage[english]{babel}

\usepackage{listings}
\usepackage{bold-extra}
\usepackage{color}
\usepackage{amsmath,amsthm} 		
\usepackage{amssymb} 			
\usepackage[flushmargin,hang]{footmisc}
\usepackage{mathrsfs,fancyhdr,latexsym}
\usepackage{graphicx}
\usepackage[normalem]{ulem}

\begin{document}

\title{Polaron mobility in oxygen-deficient and lithium doped tungsten trioxide}
\author{N. Bondarenko}
\affiliation{Division of Materials theory, Department of Physics and Astronomy, Uppsala University, Box 516, 75121 Uppsala, Sweden}
\author{O. Eriksson}
\affiliation{Division of Materials theory, Department of Physics and Astronomy, Uppsala University, Box 516, 75121 Uppsala, Sweden}
\author{N.V. Skorodumova}
\affiliation{Division of Materials theory, Department of Physics and Astronomy, Uppsala University, Box 516, 75121 Uppsala, Sweden}
\affiliation{Multiscale Materials Modelling, Department of Materials Science and Engineering,Royal Institute of Technology, SE-100 44 Stockholm, Sweden}

\begin{abstract}
Electron localization and polaron mobility in oxygen deficient as well as Li doped monoclinic tungsten trioxide have been studied. We show that small polarons formed in the presence of oxygen vacancy prefer the bipolaronic $W^{5+}- W^{5+}$ configuration whereas the $W^{6+}- W^{4+}$ configuration is found to be metastable. Our calculations suggest that bipolarons are tightly bound by the vacancy and therefore largely immobile. On the contrary, polarons formed as a result of Li intercalation can be mobile, the activation energy for polaron jumping in this case varies between 98 and 124 meV depending on the crystallographic direction.  The formation of $W^{5+}- W^{5+}$ bipolarons in $Li-WO_{3}$ is possible. When situated along $[001]$ the bipolaronic configuration is 8 meV lower in energy than two separate $W^{5+}$ polarons.  
\end{abstract}

\maketitle

\section{I. Introduction}
Phenomenon of electrochromism is found in many materials such as: organic compounds~\cite{Mortimer}, polymer composites~\cite{Carpi,Heinze} and semiconductors~\cite{Granqvist1}. One of the most studied materials of this class is tungsten trioxide due to a fortunate combination of its corrosion resistivity, sufficient light absorption and satisfactory mechanical properties that make it an attractive material for advanced applications~\cite{Hodes,Azens,Watanabe,Zheng} such as "smart windows"  technology~\cite{Svensson,Granqvist2}. 

The first mentioning of $WO_3$ coloration upon ion insertion dates back to the 19th century~\cite{Berzelius}. In the 20th century the interest to this material and its electrochromic properties revived and resulted in a large number of studies \cite{Deb,Reichman,Faughnan,Granqvist3}.  It is well known by now that "Prussian blue" coloration in the crystalline and amorphous phases of tungsten trioxide appears upon hydrogen or  alkali metal ($Li, Na$) intercalation ~\cite{Granqvist1,Svensson,Granqvist2,Zhang,Johansson1,Raj}.

Tungsten trioxide is a wide band-gap semiconductor (2.6-3.0 eV \cite{Johansson1,Granqvist4,Bamwenda})with perovskite like structure, see Fig.1. The oxide shows rich structural polymorphism \cite{Tanisaki,Diehl,Woodward,Howard}, whereby oxygen octahedrons surrounding tungsten atoms vary their orientations and shape depending both on temperature and pressure. Usually, at room temperature both ${\delta}$-$WO_3$  (triclinic $P1$ 233-290 K) and ${\gamma}$-$WO_3$ (monoclinic $P21/n$ 290-350 K) phases coexist\cite{Johansson1}. Depending on the method of preparation samples can be characterized by various ${\gamma}$/${\delta}$  ratios, micro-structure and  defectiveness. Under certain conditions a more exotic hexagonal h-${WO_3}$ phase \cite{Gerand,Khyzhun} or amorphous a-${WO_3}$ \cite{Zhang} phase can form. The structural variety of $WO_3$ offers a range of optical characteristics that is important for applications relying on light absorption \cite{Miyake,Antonaia,Sun}.

\begin{figure*}[htb]
\centering
\includegraphics[scale=0.32]{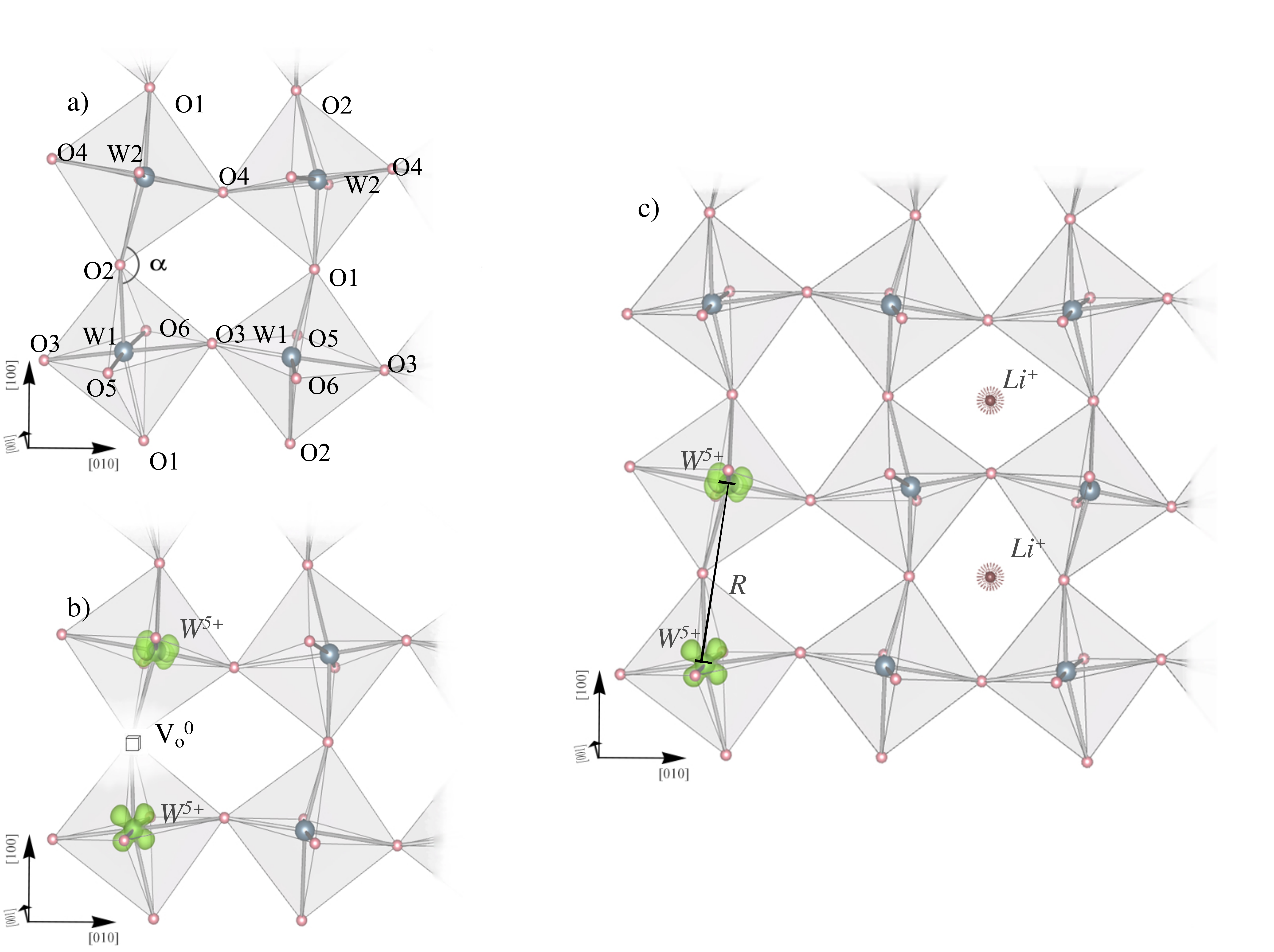}
\caption{\label{structure}  The structure of monoclinic ${\gamma}$-$WO_3$ phase without defect (a), with an oxygen vacancy (b), with two Li atoms c). Charge distributions are shown for the polaronic $W^{5+}$ states. }
\label{fig3} 
\end{figure*}

Alkali metal intercalation can also cause structural transformation of $WO_3$. For example, in the case of  $Li_{x}WO_3$ several phase transitions from monoclinic to cubic symmetry  with increasing $Li$ content have been reported ~\cite{Dey,Zhong,Lee}. It has also been shown that for $Li^+$ intercalation the degree of coloration increases proportionally to the lithium concentration whereas in the case of hydrogen insertion a saturation level of coloration is reached for a certain $H^+$ concentration ~\cite{Gerard}. Based on neutron diffraction data~\cite{Wright, Wiseman} this difference between $H$ and $Li$ has been explained by the formation of $O-H$ bonds upon hydrogen insertion.  At the same time alkali metal ions show no tendency of binding to the oxygen atoms of the oxide matrix. 

Substoichiometric $WO_{3-x}$ also shows multifarious electrochromic behavior, demonstrating different degrees of coloration depending on the concentration of oxygen vacancies ~\cite{Uppachai,Yoshimura}. Thus, experiments clearly indicate that both doping ($H,Li$) and oxygen vacancies can influence $WO_{3}$ optical properties ~\cite{Deneuville,Gerard,Pyper,Yoshimura}.  

Two general concepts of the colouration mechanism in tungsten oxide have been suggested: inter-valence charge transfer \cite{Granqvist1} and polaron models~\cite{Schrimer,Salje},  the latter being more commonly accepted. 
According to the polaron model the coloration is the result of photon absorption due to photon activated polaron hopping. A polaron can form in $WO_3$ due to electron localization at the $d$-states of tungsten, thus, turning $W^{6+}$ into $W^{5+}$ or $W^{4+}$. Light absorption can lead to the activation of the following transitions:     $W^{5+}_ {site1}+W^{6+}_ {site2} \overset{hv} \rightarrow W^{6+}_ {site1}+W^{5+}_ {site2}$ or $W^{5+}_ {site1}+W^{5+}_ {site2} \overset{hv} \rightarrow W^{6+}_ {site1}+W^{4+}_ {site2}$. We notice that more complex configurations of localized charge can form, for example, two bound polarons at neighboring sites, $W^{5+}$-$W^{5+}$, also called bipolarons (see Fig.1b,c for illustration). 

The presence of different localized charge states in $WO_3$ has been confirmed by numerous experiments~\cite{Berggren,Saenger,Hashimoto,Chen1,Johansson2,Niklarsson,Ozkan,Leftheriotis}. The formation of polarons in this oxide has also been studied theoretically by semi-empirical models~\cite{Iguchi,Stashans} and more recently by ab-initio methods employing hybrid functionals ~\cite{Wang}. More studies of the electronic and structural properties of perfect and defective $WO_3 $ can be found in Refs. ~\cite{Wijs,Migas,Heda,Chatten,Lambert}. Despite intensive research many questions regarding the properties of polarons in $WO_3 $ remain unclear. In this paper we address some of them, in particular, we study the formation and mobility of polarons in two cases: 1) in the presence of an oxygen vacancy and 2) in the presence of Li. We investigate the relative stability of polarons and bipolarons and calculate the energy barriers for polaron propagation in $WO_{3}$.

\section {II. Details of calculations}

\begin{table*}[htb]
\caption{The lattice parameters of the monoclinic cell obtained in the $HSE06$ and $DFT+U_W,U_O$ calculations together with experimental values ~\cite{Woodward}. In calculations angle ${\beta}$ was set to the experimental value $90.89^{\circ}$~\cite{Woodward}.} 
\centering 
\begin{tabular}{| p{4.0cm}  || p{4.0cm}  | p{4.0cm}  | p{4.0cm}  |}
\hline
Lattice parameters, \AA &  Experiment~\cite{Woodward} &  $HSE06$  &  $U(O_p)=9eV, U(W_d)=6eV$  \\ \hline\hline
a&7.30&7.34&7.41\\ \hline
b&7.54& 7.58&7.65\\ \hline
c&7.69& 7.73&7.81 \\ \hline
\end{tabular}
\label{tab:Tb1}
\end{table*}

The presented calculations were performed in the framework of $DFT$ using the projector augmented-wave method, as implemented in VASP ~\cite{Kresse1,Kresse2}. The $DFT+U$~\cite{Dudarev} approach using generalized gradient approximation ($GGA$) in the Perdew, Burke and Ernzerhof ($PBE$)~\cite{Perdew} parameterization was used to account for the exchange-correlation interaction. The $U$ values ($U_{O_p}=9 eV$ and $U_{W_d}=6 eV$) were determined using the linear response procedure implemented in Quantum ESPRESSO package~\cite{Giannozzi, Cococcioni}. Hereafter we refer to this set of calculations as to $DTF+U_W,U_O$. We also performed the $DFT+U$ calculations with $Hubbard-U$ applied only to the d-states of $W (DFT+U_W$). Additionally, we used the screened hybrid functional of Heyd, Scuseria, and Ernzerhof $(HSE06)$~\cite{Heyd1,Heyd2} with the Coulomb potential separation parameter, ${\omega}$, of $0.2~{\AA}^{-1}$~and the $HF$ mixing constant of 0.25.
 
Our $DFT+U$ calculations were performed for the 2x2x2 unit cell (192 $O$ and 64 $W$) and 2x1x1 unit cell (48 $O$ and 16 $W$), both using a $\Gamma$-point centered 2x2x2 k-point grid for  the Brillouin zone integration. The 2x1x1 unit cell was also used in the hybrid calculations. The following states were treated as valence states: $5p^6, 5d^4, 6s^2 (W)$ and $2s^2, 3p^4 (O)$ and $1s^1, 2p^0 (Li)$.  All the calculations were spin-polarized. The cut-off energy was 650 eV.  

The structure relaxation was done using the following scheme.  First, we setup an initial supercell using the experimental parameters~\cite{Woodward} then we relaxed the structure in two steps: 1) first we fixed the atomic positions and relaxed only the cell parameters; 2) next we fixed the cell volume and relaxed the atomic positions. The latter were relaxed until the Hellmann-Feynman forces acting on atoms became smaller than $10^{-4}$ eV/\AA.
We repeated this two step relaxation routine a few times to ensure a stable result.  We notice that during the relaxation we kept the symmetry of the lattice monoclinic.

Here we study only uncharged oxygen vacancies, for which the vacancy formation energy, $E_{v}$, can be calculated as
\begin{equation}
\begin{split}
\label{HSE}
      E_{v}=E_{W_xO_{3x}}-E_{W_xO_{3x-1}}-1/2E_{O_2},
\end{split}
\end{equation}

The barriers for polaron transitions were calculated using both the linear interpolation scheme $(LIS)$ as in our previous work ~\cite{Bondarenko} and the nudged elastic band method $(NEB)$~\cite{Henkelman}. $NEB$ is an efficient method for studying atomic and molecular diffusion, phase transitions, chemical reactions as well as polaronic transitions in oxides ~\cite{Henkelman, Sheppard, Gonzalez, Xie, Ong}. We used 10 intermediate configurations in LIS and 5 images for NEB. 
\begin{table*}[htb]
\caption{Bond distances and angles obtained for the monoclinic $WO_3$ structure using $DFT+U_{W},U_{O}$. The corresponding experimental values are shown in parentheses (~\cite{Woodward}). The oxygen position notations are as shown in Fig. 1a. The subscripts x,y,z are added for convenience to indicate the orientation of  the corresponding $W^{5+}$-$W^{5+}$ pair in the crystal.}
\centering 
\begin{tabular}{| p{3.0cm}  || p{3.0cm}  | p{3.0cm}  | p{3.0cm}  |}
\hline
Oxygen position&${\alpha}$ , $ (^{\circ})$ & WO-bond1, ($\AA$)&WO-bond2, ($\AA$)\\ \hline
$O1_x$, [100]&160.1, (154.8)&1.90, (1.89)&1.89, (1.81)\\ \hline
$O2_x$, [100]&160.7, (155.7)&1.90, (1.95)&1.89, (1.91)\\ \hline
$O3_y$, [010]&163.6, (161.5)&1.84, (1.82)&2.03, (2.02)\\ \hline
$O4_y$, [010]&163.4, (160.1)&2.04, (2.06)&1.84, (1.78)\\ \hline
$O5_z$, [001]&170.5, (164.2)&2.10, (2.16)&1.80, (1.71)\\ \hline
$O6_z$, [001]&169.7, (157.3)&1.81, (1.81)&2.09, (2.11)\\ \hline
\end{tabular}
\label{tab:Tb1}
\end{table*}
%
\section{III. Results and discussion}
\subsection{A.	Electron localization in oxygen deficient $\gamma-WO_3$}

\begin{figure}[htb]
\begin{center}
\includegraphics[scale=0.40]{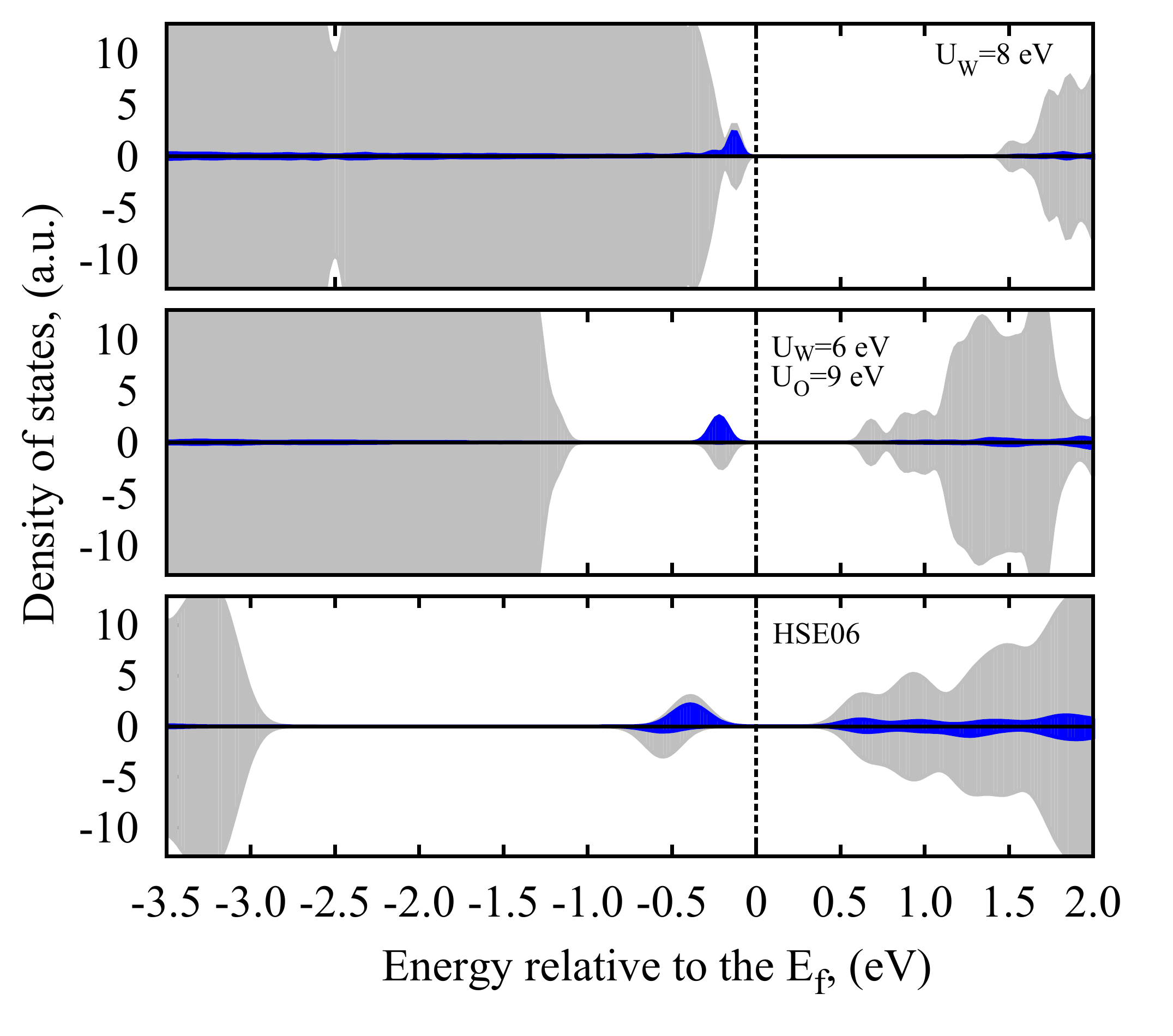}
\end{center}
\caption{\label{structure}  $DOSs$ calculated for the 2x1x1 supercell with an oxygen vacancy and associated $W^{5+}-W^{5+}$ bipolaron(see Fig.1b).  The top, middle and bottom panels present $DOSs$ obtained with $DFT+U_W$, $DFT+U_W,U_O$, $HSE06$, respectively. The total $DOSs$ are shown in grey, the polaronic $W_d$ states are in blue.}
\label{fig5} 
\end{figure}

To properly model small polarons one needs a method adequately describing the on-site charge localization and lattice distortions around this site ~\cite{Andersson,Arapan,Bondarenko,Droghetti,Maxisch}. Electron localization in correlated $d-$ and $f-$metal oxides usually cannot be correctly described by standard $DFT$ approximations, which due to the presence of the self-interaction term tend to overestimate the stability of the delocalized solution.  The Coulomb interactions can be corrected, for example, by means of the $Hubbard-U$ parameter \cite{Anisimov1}. It can be treated as an adjustable parameter to better describe some experimental data~\cite{Andersson} or it can be obtained from restricted ab initio calculations \cite{Cococcioni}. Traditionally in oxides $Hubbard-U$ is applied to the $d-$ or $f-$states of cations, in some cases, however, the description of oxide property can be further improved by applying $U$ also to the $p-$states of oxygen \cite{Deskins, Ma, Erhart}. 
Another way to cure the self-interaction problem of standard $DFT$ is to employ the so-called hybrid functionals, where a certain amount of nonlocal
Fock exchange is added \cite{Becke}. Both approaches have successfully been used to describe electron localization or small polaron formation in different oxides ~\cite{Andersson,Arapan, Bondarenko,Droghetti,Maxisch,Wang,Franchini}. Hybrid functionals are more computationally demanding than $DFT+U$ and therefore their usage is limited to rather small unit cells.

To find the optimal functional for the description of polarons in $WO_3$ we have performed three sets of calculations using the 2x1x1 unit cell (64 atoms), namely, $DFT+U_W$, $ DFT+U_{W},U_{O}$ and $ HSE06$ (see Sec. II for details). As a test case we considered an oxygen vacancy with two unpaired electrons localized on both sides of the vacancy in the $W^{5+}-W^{5+}$  configuration (Fig.1b).

First, within the $DFT+U_W$ approach we varied $U_{W}$ from $2$ to $12~eV$ in the manner described in Ref.\cite{Bondarenko}. We analyzed the localization patterns for different values and found that for $U_{eff} = 8 eV$ and higher almost precisely one electron was localized at the d-orbitals of each of the two $W^{5+}$ tungsten atoms. The local magnetic moment (${\mu}_{W}$) for each $W$ in this case was  $0.96~{\mu}B$. 
Next, we performed calculations using $DFT+U_{W},U_{O}$. In this case the local magnetic moment due to the electron localization at the $W_d$-states was $0.86~{\mu}B$.
Finally, our $HSE06$ calculation resulted in ${\mu}_{W}=0.64~{\mu}B$.

The density of states ($DOSs$) obtained in the three types of calculations are compared in Fig.2. The band gaps are expectedly underestimated by both versions of the $DFT+U$ calculations: $E_g(DFT+U_W$)=1.43 eV, $E_g(DFT+U_{W},U_{O})$= 1.64 eV that should be compared to the experimental value of 2.8  eV ~\cite{Johansson1}. At the same time, $HSE06$ overestimates the gap: $E_g(HSE06)=3.19$ eV.  
In the case of $DFT+U_W$ the localized polaronic d-states are situated at the upper edge of the valence band whereas for $HSE06$ and $DFT+U_{W},U_{O}$ these states are, respectively, 0.8 eV  and 2.5 eV above the valence band top. The position of the localized states with respect to the conduction band edge is similar for both approximations. We notice that the position of the polaronic peak in the gap obtained by $HSE06$ is in good agreement with previously reported $B3LYP$ results ~\cite{Wang}.  

The magnetic state of the obtained bipolaron is degenerate according to $HSE06$ whereas in the case of $DFT+U_{W},U_{O}$ the antiferromagnetic coupling of local spins is favored by about 20 meV over the ferromagnetic one.
Using $DFT+U_{W},U_{O}$ we also examined $DOSs$ calculated for $Li-WO_3$ using the 2x2x2 unit cell and found them to be very similar to those calculated for the vacancy case. The only noticeable difference was a 40 meV shift of the polaronic peak down in energy. 

The lattice parameters obtained using $DFT+U_{W},U_{O}$ and $HSE06$ together with experimental data are shown in Table I. They demonstrate reasonable agreement with the $HSE06$ values being a bit closer to the experimental parameters. Based on the analysis of $DOSs$, localization behavior and structures obtained in the three sets of calculations we conclude that the $DFT+U_{W},U_{O}$ approach is the most appropriate for studying polarons in $WO_3$ as it shows good agreement with the results of $HSE06$ and, at the same time, allows us to treat large supercells. Therefore, further on we study polaron mobility in $WO_3$ using the $DFT+U_{W},U_{O}$ approach.    

\subsection{B. Analysis of oxygen vacancy positions and vacancy trapped polarons.}

\begin{table*}[htb]
\caption{Vacancy formation energies, $E_v$, and relevant W-W distances, $R$,  for non-equivalent oxygen sites and $W^{5+}-W^{5+}$~(see Fig.1c) and $W^{6+}-W^{4+}$  configurations. The numbers in parentheses show the corresponding $W-W$ distances in the perfect $WO_3$ cell.}
\centering 
\begin{tabular}{| p{3.0cm}  || p{3.0cm}  | p{3.0cm}  | p{3.0cm}  |p{3.0cm}  |}
\hline
Oxygen crystallographic positions&$E_v (eV), W^{5+}-W^{5+}$&$E_v (eV), W^{6+}-W^{4+}$&$R (\AA), W^{5+}-W^{5+}$&$R (\AA), W^{6+}-W^{4+}$\\ \hline
$O1_x$,&2.79&2.90&3.94, (3.72)&3.87\\ \hline
$O2_x$,&2.77&2.86&3.92, (3.74)&3.86\\ \hline
$O3_y$,&2.66&2.78&4.01, (3.83)&3.90\\ \hline
$O4_y$,&2.62&2.73&4.05, (3.84)&3.92\\ \hline
$O5_z$,& \textbf{2.36} &2.51&\textbf{4.35, (3.91)}&4.19\\ \hline
$O6_z$,&2.40&2.54&4.31, (3.89)&4.14\\ \hline
\end{tabular}
\label{tab:Tb1}
\end{table*}

 The monoclinic phase of $WO_3$ ($2c_{1}/n$) has two tungsten and six oxygen non-equivalent atoms  ~\cite{Woodward}, as illustrated in Fig.1a. The structure is rather complex, each $O$ atom has two neighboring $W$ atoms but the $W-O-W$ chains differ in all the directions. Along [100] we find $...-O1-W1-O2-W2-O1-...$ and $...-O2-W1-O1-W2-O2-...$ chains, along [001] $...-O6-W1-O5-W2-O6-...$ and $...-O5-W1-O6-W2-O5-...$ chains interconnected in the [010] direction by alternating $...W1-O3-W1-O3-W1-...$ and $...-W2-O4-W2-O4-W2-...$ chains (for notations see Fig.1a).  The $W-O$ bonds in all the directions have alternating shorter and longer distances and slightly different $O-W-O$ angles, which are shown in Table II, both calculated with $DFT+U_{W},U_{O}$ and from the experimental data ~\cite{Woodward}. 
 
 \begin{figure}[htb]
\begin{center}
\includegraphics[scale=0.36]{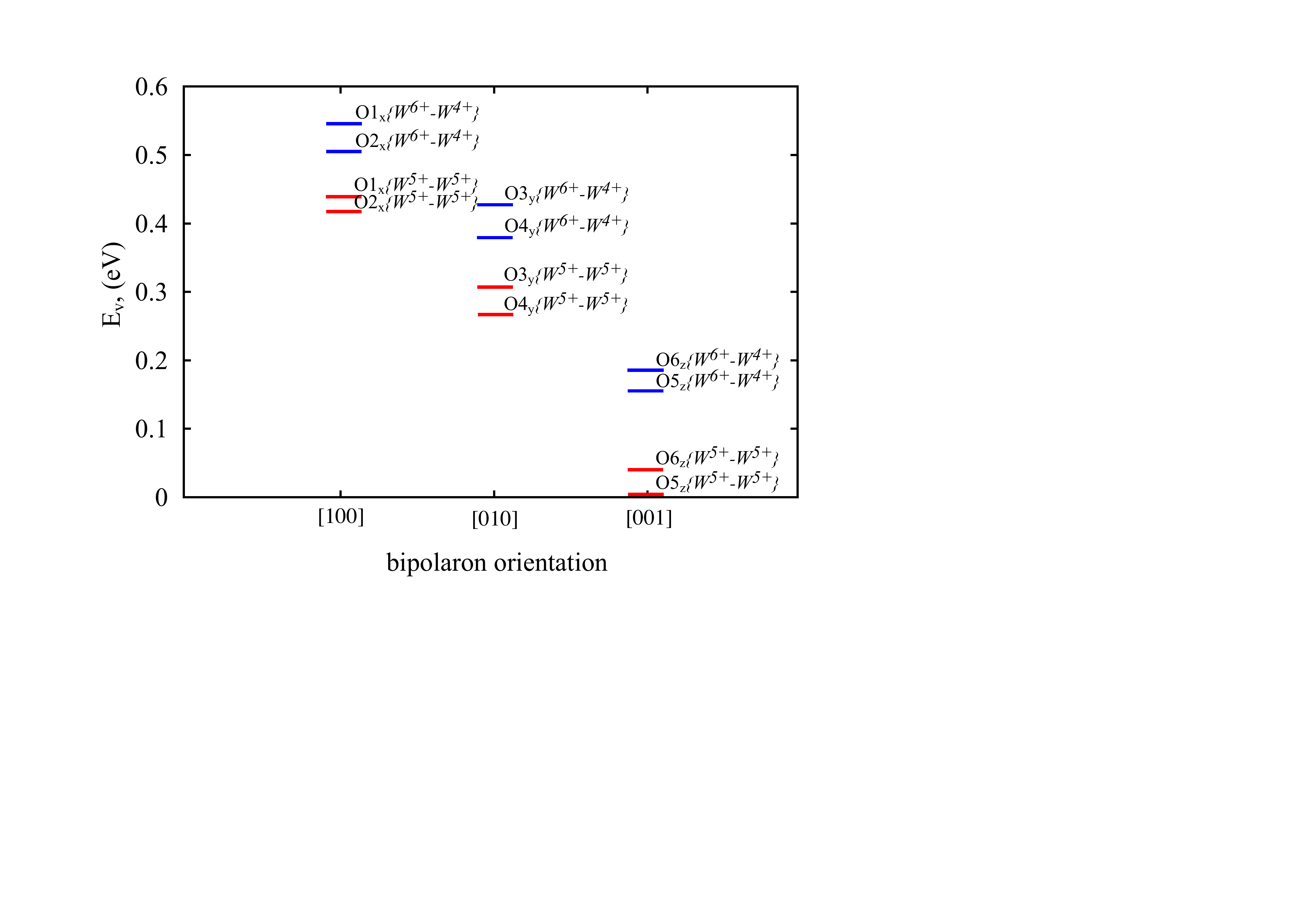}
\end{center}
\caption{\label{structure} Vacancy formation energies for different vacancy positions and two electronic configurations: $W^{6+}-W^{4+}$ and $W^{5+}-W^{5+}$. The energies are shown with respect to the lowest formation energy, the one for the $O5_z$ position, with the energy of 2.36 eV. See also Table III.}
\label{fig5} 
\end{figure}

\begin{figure}[htb]
\centering
\includegraphics[scale=0.35]{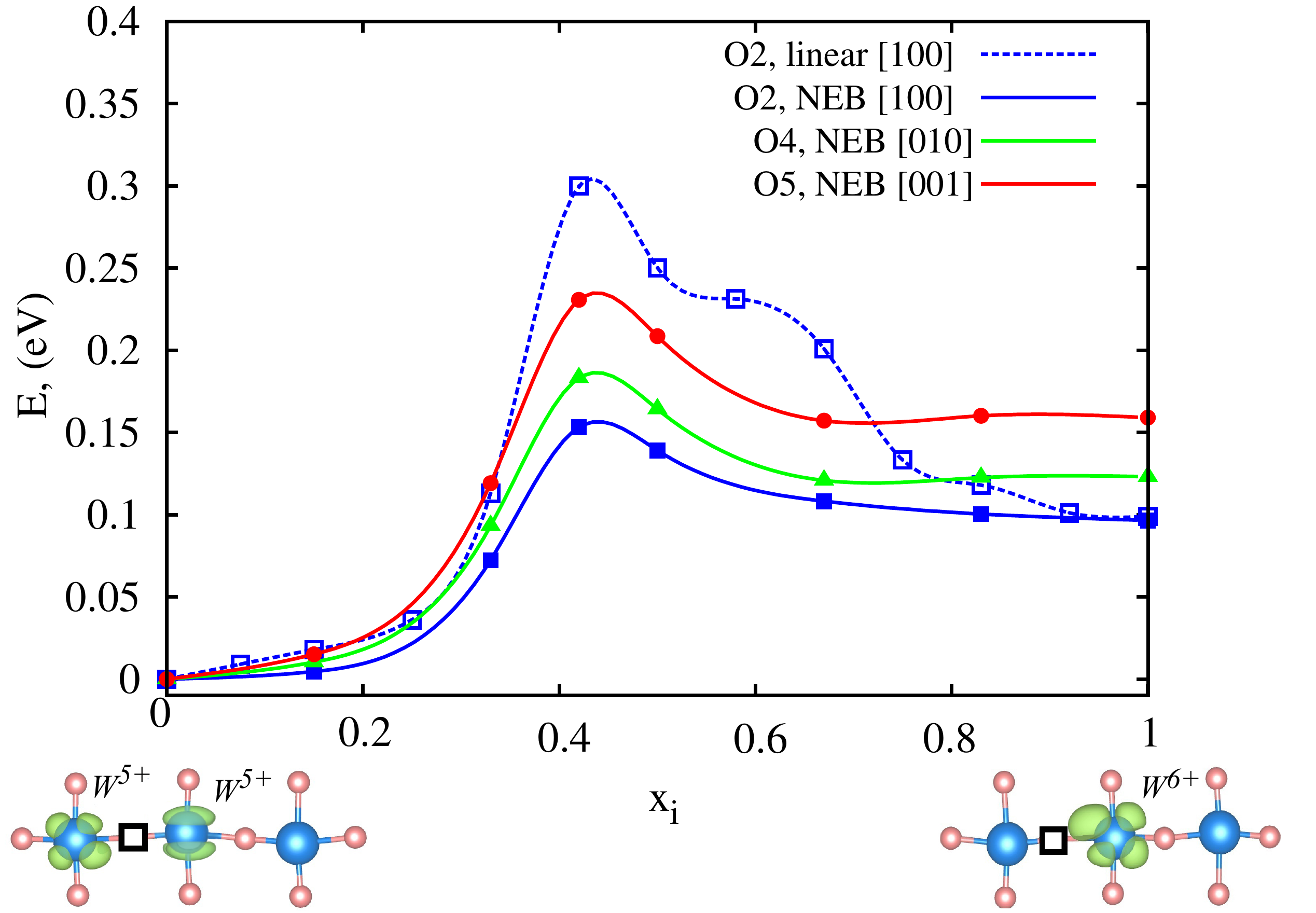}
\caption{\label{structure}  Energies along the transition path between the $W^{5+}-W^{5+}$ and $W^{6+}-W^{4+}$ configurations for three vacancies ($O2_x$, $O4_y$ and $O5_z$) as estimated by $NEB$. The energies are given with respect to that of $W^{5+}-W^{5+}$ in the corresponding direction. For comparison for $O5_z$ we show the results obtained by the linear interpolation method $(LIS)$ \cite{Bondarenko}. Charge distributions corresponding to the initial x=0 ($W^{5+}-W^{5+}$) and final x=1 ($W^{4+}$)  transition points are shown for illustration.}
\label{fig6} 
\end{figure}

Using the structure obtained in $DFT+U_{W},U_{O}$ we calculated the vacancy formation energetics for the six oxygen positions (see Fig.1a) and two bipolaronic configurations:  $W^{6+}-W^{4+}$ and  $W^{5+}-W^{5+}$ bipolaron (see Fig.1b). Both electronic configurations could be stabilized for all the vacancies. The results are presented in Fig.3 and Table III. The lowest vacancy formation energy, 2.36 eV, is found for the oxygen vacancy at site $O5_z$ with bipolaronic configuration $W^{5+}-W^{5+}$ and the highest energy, 2.90 eV, for the $O1$ position in the $W^{6+}-W^{4+}$ configuration. The formation energies for the $W^{6+}-W^{4+}$ configurations are systematically higher than those for the $W^{5+}-W^{5+}$ ones (Table III). 
The vacancy formation energy, $E_{v}$,  depends on the orientation of the bipolaron in the lattice, $E_{v}$ averaged for each direction can be sorted in the following order: $E_{v}[100]>E_{v}[010]>E_{v}[001]$. These results are in agreement with previous studies carried out by standard $DFT$ ~\cite{Lambert}, where the oxygen atoms with the $W-O$ bonds along the $[001]$ direction were indicated as favorable sites for vacancy formation. Using the $B3LYP$ hybrid functional Wang et al. ~\cite{Wang} also reported the formation of the $W^{5+}-W^{5+}$ configuration aligned with $[001]$.

The data in Table III show that the vacancy formation energy decreases as the $W-W$ distances increase and lowest $E_v$ corresponds to the largest $W-W$ separation. The formation of bipolarons ($W^{5+}-W^{5+}$) leads to a noticeable increase in the $W-W$ bond lengths due to Coulomb repulsion. These local lattice distortions are, however, different for different crystallographic directions. In the [100] and [010] directions the $W-W$ distance increase is of about 0.20~\AA~whereas for the $O6_z$ and $O5_z$ vacancy positions ($[001]$) it is 0.42~\AA~and 0.44~\AA,~respectively. We see a similar trend for the $W^{6+}-W^{4+}$ polaronic configurations (Table III), however, the changes of the $W-W$ distance are smaller in this case. These results indicate certain softening of phonon modes along $[001]$. They also indicate that in the case of monoclinic $WO_3$ the electronic contribution to the energy of vacancy formation dominates over the elastic contribution.

Our calculations suggest that the bipolaronic states associated with an oxygen vacancy are tightly trapped as all our attempts to localize charge away from the vacancy were unsuccessful. Therefore, we can expect that in the presence of oxygen vacancies bipolarons are immobilized in the lattice, however, the transitions between vacancy bound $W^{5+}-W^{5+}$ and $W^{6+}-W^{4+}$ could still contribute to the coloration effect in $WO_3$.  To study the transition between these two configurations we chose oxygen vacancies with lowest $E_v$ in each crystallographic direction, namely, $O2_x$, $O4_y$ and $O5_z$. Our results, presented in Fig.4, show that the transition barriers vary from 150 to 230 meV, being the smallest for the $[100]$ direction and largest for $[001]$. This result is in agreement with similar trends shown by the energy differences between the $W^{5+}-W^{5+}$ and $W^{6+}-W^{4+}$ configurations: $O2_x,~(90 ~meV) < O4_y,~(110~meV) < O5_z,~(150~meV)$ and the $W-W$ bond length differences: $O2_x,~(0.06~\AA)< O4_y,~(0.13~\AA)< O5_z,~(0.16~\AA)$.   We also notice that the $NEB$ relaxation results in lower barriers compared to those obtained with $LIS $ (Fig.4). This is understandable as $NEB$ allows one to optimize structures along the transition path whereas $LIS$ uses static configurations.

Lattice distortions around polarons and especially bipolarons can be quite substantial reaching far beyond first coordination shells. In the case studied here we find that the local distortions around the $bipolaron-vacancy$ complex spreads as far as 10-11~\AA~in the direction of the $W^{5+}-W^{5+}$ pair alignment. The 2x2x2 cell used in our calculations is big enough to accommodate such a defect. Our results, however, highlight the importance of using large unit cells in modeling polarons and bipolarons at least if one wishes to study isolated quasiparticles of the sort.

\subsection{C. Polaron formation and transitions in $Li$ doped monoclinic $\gamma$ phase.}

Polarons can also form in the presence of impurities, dopants or due to electron injection. Here we report our results of the polaron formation and mobility in $Li$ doped monoclinic $WO_3$. First we consider one lithium atom situated in an octapore in the supercell of 256 atoms, i.e. $WO_3Li_{0.016}$. This is a realistic concentration often studied in experiment \cite{Larsson}. After the relaxation Li stayed close to the octapore center (Fig 1c), no bond formation with the surrounding oxygen atoms was detected. It was possible to localize $W^{5+}$ polaron at different tungsten sites. The association energy of $Li^+$ and $W^{5+}$ is of the order of 30 meV, which was estimated by calculating the energies of $Li^{+}-W^{5+}$ separated by 3.92~\AA~and 7.71~\AA. 

Further, to study polaron mobility we used tungsten sites away from $Li$ ($\sim$ 6-8~\AA)~simulating a free polaron propagation in the $WO_3$ matrix. Fig.5 shows the energy barriers for polaron jumping between two neighboring W sites in the three crystallographic directions. As expected the $LIS$ barriers are again higher than those obtained using $NEB$. The $NEB$ barriers vary from 98 for $[001]$ to 124 meV for $[100]$. These values are in surprisingly close agreement with the activation energy for polaron hopping estimated from basic models using experimental information, 180 meV~\cite{Larsson}. 

If the concentration of lithium ions and polarons increases the number of possible configurations of localized charges becomes larger. We study the possibility of a bipolaron formation by considering two Li atoms in the 256 atom $WO_3$ supercell ($WO_3Li_{0.032}$). Next to the bipolaronic $W^{5+}-W^{5+}$ configurations we also consider the formation of two separate $W^{5+}$ polarons and one $W^{4+}$. 
In Fig.6 we show the energies for different electronic configurations and transitions between them obtained using $NEB$. We start from the situation when both electrons donated by the two $Li$ atoms are delocalized in the oxide matrix (x=0, Fig.6), then we localize one $W^{5+}$ polaron that happens without any barrier (x=2, Fig.6). After that we localize the second electron at the next nearest $(NN)$ $W$ site from first $W^{5+}$, which neither involves any barrier(x=3, Fig.6).
The energies of such $NN$ configurations are very close to each other for all the orientations (Fig.6). Next we model a transition of one $W^{5+}$ from the $NN$ site to the nearest $(N)$ $W$ site, thus, forming a bipolaron along the $[100]$, $[010]$ and $[001]$ directions (x=3 $\to$ x=5, Fig.6). This transition involves a barrier of about 90 meV. The resulting bipolaron situated along $[001]$ has the lowest energy not only among the bipolarons but among all the considered configurations (Fig.6). The $[001]$ bipolaron is 8 meV lower in energy than the $[001]$ $NN$ configuration. At the same time, the energies of bipolarons oriented along $[100]$ and $[010]$ are about 20 meV higher than those of the $NN$ configurations. 
We notice that the localization of two electrons at one tungsten site ($W^{4+}$) is energetically unfavorable as this state lies more than 300 meV above any $W^{5+}-W^{5+}$ configuration. Therefore, our results suggest that in $Li$ doped $WO_3$ one can expect the formation of $W^{5+}$ polarons and $W^{5+}-W^{5+}$ bipolarons whereas the presence of the $W^{4+}$ state, often detected in experiments ~\cite{Ozkan,Johansson2,Uppachai,Gerard,Sun,Leftheriotis}, most likely can be attributed to oxygen vacancies, around which the $W^{4+}$ state is found to be metastable.

\begin{figure}[htb]
\centering
\includegraphics[scale=0.35]{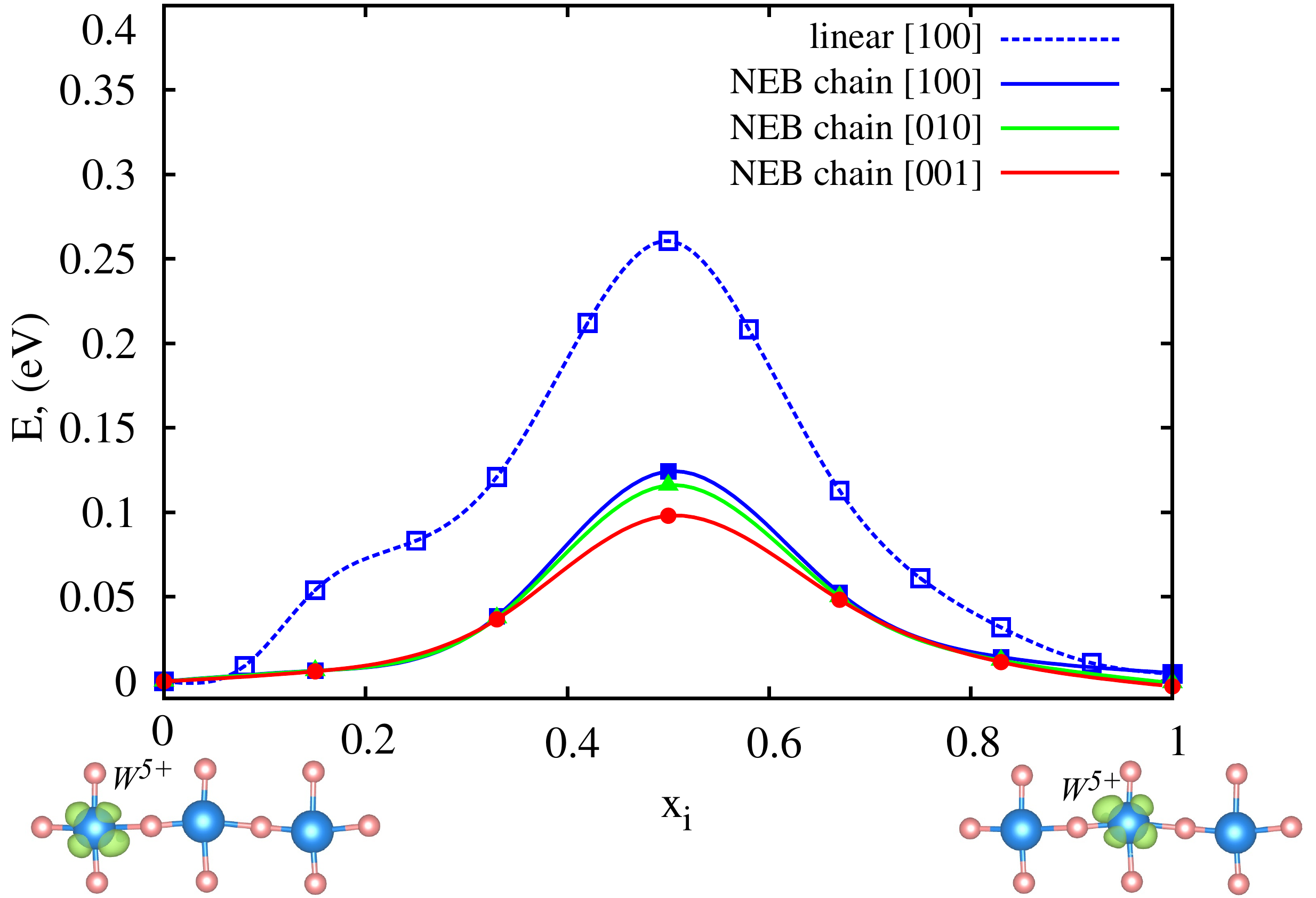}
\caption{\label{structure} Energies along the transition paths of $W^{5+}$ from one tungsten site to the other in the case of $Li$ doped $\gamma$ phase. The energies are given with respect to the energy of $W^{5+}$. The solid lines show the $NEB$ energies for all the three crystallographic directions. For comparison, for $[100]$ we also show the energies obtained in $LIS$ calculations (dashed line). Charge distributions for $W^{5+}$ at the initial $x=0$ and final $x=1$ transition points are shown for illustration.}
\label{fig6} 
\end{figure} 
\begin{figure*}[htb]
\centering
\includegraphics[scale=0.50]{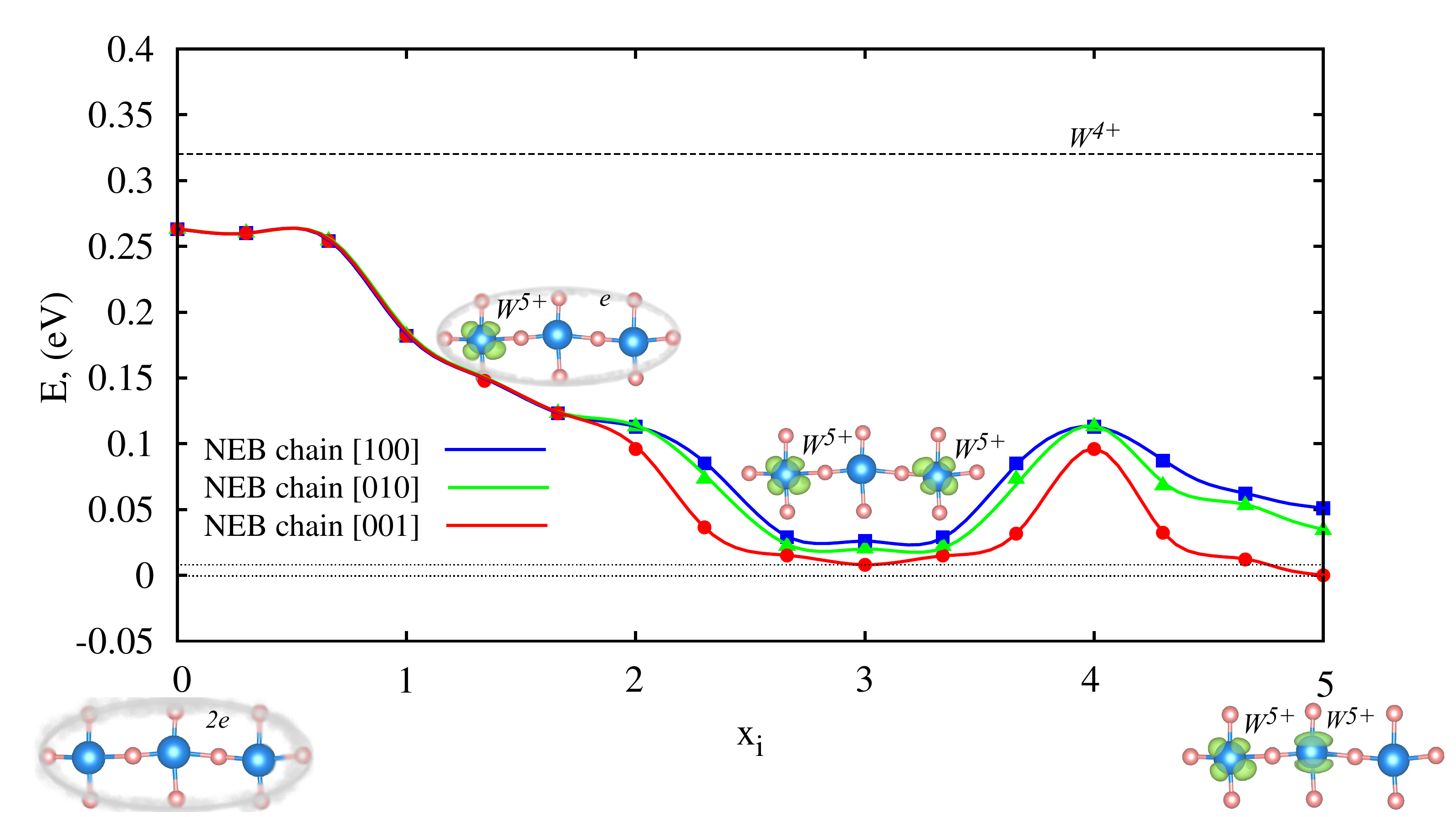}
	\caption{\label{structure}Energies along the transition paths calculated using $NEB$ for the case of two intercalated $Li$ atoms (Fig.1c). The energies are shown with respect to the lowest energy configuration $W^{5+}-W^{5+}$ situated along [001]. The following polaronic transitions are shown: from x=0 (two delocalized electrons) $\to$ x=2 (one electron localized as $W^{5+}$, the other is delocalized) then $\to$ x=3 (two $W^{5+}$ as next nearest neighbors) and further $\to$ x=5 ($W^{5+}- W^{5+}$ bipolaron configurations). These transitions are calculated along $[100], [010]$ and $[001]$. The energy of the $W^{4+}$ configuration is marked by the dashed line. Two dotted lines indicate the energy difference between the $W^{5+}-W^{5+}$ bipolaron and $NN$ configuration  situated along $[001]$ direction. Charge distributions for the $W^{5+}$ polarons at x=2, 3 and 5 are shown for illustration.}
\label{fig7} 
\end{figure*} 
%

\section {IV.Conclusion}

We have studied the applicability of $HSE06$ and two $DFT+U$ approaches to modelling of polarons in $WO_3$. We have shown that the $DFT+U$ approach with two $Hubbard-U$ parameters determined and applied simultaneously to the d-W ($U_W$=6 eV) and p-O ($U_O$=9 eV) states provides a proper description of polaron formation and transitions in $WO_3$. At the same time, this method allows us to calculate large unit cells necessary for adequate modelling of polarons and bipolarons. Using this approach we have studied the vacancy energetics for the six non-equivalent oxygen positions and two electronic configurations:  $W^{5+}-W^{5+}$ and $W^{6+}-W^{4+}$. The $W^{5+}-W^{5+}$ bipolarons situated along $[001]$ are the most favorable electronic configurations around vacancies. We have also studied polaron and bipolaron formation and energetics for Li doped $WO_3$. In this case the $W^{5+}-W^{5+}$  bipolaronic configuration aligned with $[001]$ has again the lowest energy but winning only 8 meV over two separated $W^{5+}$. Our results suggest that the polarons formed due to oxygen vacancies are immobile, at the same time, the $W^{4+}$ state is metastable and the $W^{5+}-W^{5+}$ to $W^{6+}-W^{4+}$ transition is possible with a barrier of 150 meV. On the contrary, polarons formed in $Li$ doped tungsten oxide are mobile with the minimum activation energy (98 meV) in the [001] direction.  The $W^{4+}$ state is 300 meV higher in energy than any studied $W^{5+}$ configuration and, therefore,  $W^{4+}$ is unlikely to form in perfect $Li-WO_3$ without presence vacancies or similar structural defects. 

\section {V. Acknowledgments:}
We acknowledge the financial support by the Swedish Energy Agency (Energimyndigheten, STEM) eSSENCE, STANDUPP and the Swedish Research Council (Vetenskapsr\aa det).  Supercomputer time was granted by the Swedish National Infrastructure for Computing (SNIC). N.B. is grateful to National Supercomputer Centre (NSC) for provided support. O.E. Acknowledges in addition the KAW. Authors are thankful to C-G. Granqvist and G. Niklasson for valuable discussions.

\end{document}